\documentstyle[11pt]{article}

\makeatletter
\renewcommand{\section}{\@startsection{section}{1}{0pt}%
        {-3.5ex plus -1ex minus -.2ex}{2.3ex plus .2ex}%
        {\large\bf\protect\raggedright}}

\renewcommand{\subsection}{\@startsection{subsection}{2}{0pt}%
        {-3ex plus -1ex minus -.2ex}{1.4ex plus .2ex}%
        {\normalsize\bf\protect\raggedright}}

\renewcommand{\@oddhead}{\raisebox{0pt}[\headheight][0pt]{%
   \vbox{\hbox to\textwidth{\rightmark \hfil \rm \thepage \strut}\hrule}}}
\renewcommand{\@evenhead}{\raisebox{0pt}[\headheight][0pt]{%
   \vbox{\hbox to\textwidth{\thepage \hfil \leftmark \strut}\hrule}}}

\makeatother

\newcommand{\sect}[1]{Sec.\,#1}

\def\nqq{\hspace{-2em}}
\def\nhq{\hspace{-0.5em}}

\def\cm{\hspace{1cm}}
\def\inch{\hspace{1in}}

\def\eq{Eq.\,}

\def\beq{\begin{equation}}
\def\eeq{\end{equation}}
\def\bear{\begin{eqnarray}}
\def\al{&\nhq}
\def\lal{&&\nqq {}}               
\def\bearr{\begin{eqnarray} \lal}
\def\ear{\end{eqnarray}}
\def\earn{\nonumber \end{eqnarray}}

\def\nn{\nonumber\\ {}}

\def\nnn{\nonumber\\ \lal }

\def\eql{\al =\al}

\def\e{{\,\rm e}}

\def\sign{\mathop{\rm sign}\nolimits}

\def\const{{\rm const}}

\def\then{\ \Rightarrow\ }

\newcommand{\vars}[1]{\left\{\begin{array}{ll}#1\end{array}\right.}

\def\sph{spherically symmetric\ }

\def\bh{black hole}
\def\bhs{black holes}
\def\TH{T_{\rm H}}

\begin{document}
\thispagestyle{empty}
\rightline{\bf gr-qc/9710092}
\medskip

\begin{center}
{\Large\bf COLD BLACK HOLES \\[5pt] IN SCALAR-TENSOR THEORIES}
\bigskip

{\bf K. A. Bronnikov\footnote{e-mail: kb@cce.ufes.br;
permanent address: Centre for Gravitation and Fundamental Metrology,
VNIIMS, 3-1 M. Ulyanovoy St., Moscow 117313, Russia, e-mail:
kb@goga.mainet.msk.su}
C. P.
Constantinidis\footnote{e-mail: clisthen@cce.ufes.br}, R. L.
Evangelista\footnote{e-mail: rleone@cce.ufes.br} and J. C.
Fabris\footnote{e-mail: fabris@cce.ufes.br}} \medskip

Departamento de F\'{\i}sica, Universidade Federal do Esp\'{\i}rito Santo,
Vit\'oria, Esp\'{\i}rito Santo, Brazil \medskip

\end{center}

\begin{abstract}
We study the possible existence of black holes in scalar-tensor theories of
gravity in four dimensions. Their existence is verified for anomalous
versions of these theories, with a negative kinetic term in the Lagrangian.
The Hawking temperature $\TH$ of these holes is zero, while the horizon
area is (in most cases) infinite. It is shown that an infinite value
of $\TH$ can occur only at a curvature singularity rather than a
horizon.  As a special case, the Brans-Dicke theory is studied in more
detail, and two kinds of infinite-area \bhs\ are revealed, with finite and
infinite proper time needed for an infalling particle to reach the horizon.
\end{abstract}

\section{Introduction}

This study was to a certain extent stimulated by a controversy in the
recent literature:  the paper by Campanelli and Lousto \cite{lousto}
asserts that in the well-known family of static, \sph vacuum solutions of
the Brans-Dicke theory there exists a subfamily which possesses all
properties of \bh\ solutions, but (i) these solutions exist only for
negative values of the coupling constant $\omega$ and (ii) the horizons
have an infinite area.  These authors argue that large negative $\omega$
are compatible with modern observations and that such \bhs\ may be of
astrophysical relevance. On the other hand, H. Kim and Y. Kim
\cite{kim}, agreeing that there are non-Schwarzschild \bhs\ in the
Brans-Dicke theory, claim that such \bhs\ have unacceptable thermodynamical
and geometric properties and are therefore physically irrelevant;
meanwhile, they ascribe such solutions to positive values of $\omega$.

The aim of this work is not only to make the situation clear, but a bit
wider: to reveal possible \bh\ solutions among static, spherically
symmetric solutions of the general (Bergmann-Wagoner) class of
scalar-tensor theories of gravity, which may be described in terms of
the function $\omega(\phi)$; the Brans-Dicke theory ($\omega=\const$)
will be used just as an example.  One of the reasons for such an approach
is that, by modern views, it is rather probable that this coupling
parameter could have been sufficienly small and could appreciably affect
the physical processes in the early Universe, but by now became large,
making the theory very close to general relativity in observational
predictions.

To identify the presence of a regular horizon, we use such criteria as
the finiteness of the Kretschmann scalar and the Hawking temperature.
They have been used in \cite{lousto,kim} as well, but we try to
perform a more complete analysis. In particular, we prove that an infinite
Hawking temperature implies the divergence of the Kretschmann scalar.

The case of the Brans-Dicke theory is studied in more detail. We find that
regular event horizons do exist, but those with an infinite area (we call
them Type B horizons), for negative values of $\omega+3/2$, thus confirming
the conclusions of \cite{lousto}.

Moreover, we show that black holes can exist in other
theories, where the coupling $\omega\ne\const$. In all such cases,
the Hawking temperature is also zero (cold \bhs) and the horizon area must
be infinite in most cases (except $k<0$, see \sect 3).

\section{The Kretschmann Scalar and the Hawking Temperature}

We would like to begin with some general consideration of static, \sph
space-times, with results to be used in the subsequent sections.

The general form of the metric of such space-times is
\beq                                                         \label{m1}
   ds^2 = \e^{2\gamma}dt^2 - \e^{2\alpha}du^2 - \e^{2\beta}d\Omega^2 \, ,
\eeq
where $\gamma$, $\alpha$ and $\beta$ are functions of $u$ only
and $d\Omega^2= d\theta^2 + \sin^2\theta d\phi^2$.

Event horizons, to be discussed below, must be regular,
hence all curvature invariants must be finite there.
The finiteness of the Kretschmann invariant
$R^{\mu\nu\lambda\gamma}R_{\mu\nu\lambda\gamma}$ is known to be the most
efficient criterion of this type.  We will, however, prove another,
sometimes more convenient criterion:  an infinite Hawking
temperature indicates that an assumed horizon exhibits a singularity (in
fact, in this case the use of the term ``horizon" is doubtful).

The Kretschamnn scalar for the metric (\ref{m1}) may be written as
\beq                                                        \label{k}
	K = 4K_1^2 + 8K_2^2 + 8K_3^2 + 4K_4^2 \, ,
\eeq
where
\bear
K_1 &=& {R^{01}}_{01} = - \e^{-\alpha - \gamma}
         \biggr(\gamma'\e^{\gamma - \alpha}\biggl)'  ,\nn
K_2 &=& {R^{02}}_{02} = {R^{03}}_{03} = - \e^{-2\alpha}\beta'\gamma' ,\nn
K_3 &=& {R^{12}}_{12} = {R^{13}}_{13} = - \e^{-\alpha - \beta}\biggr(\beta'
         \e^{\beta-\alpha}\biggl)' \, , \nn
K_4 &=& {R^{23}}_{23} = \e^{-2\beta} - \e^{-2\alpha}{\beta'}^2
\ear
where a prime denotes $d/du$.
The structure of \eq(\ref{k}) indicates that an infinite value of any
$K_i$ implies the presence of a singularity at a given point of the
space-time.

Let us now prove the following statement:
\begin{itemize}
\item
If, at a certain surface $u = u^* $ of a static, spherically symmetric
configuration, $g_{00}= 0$ (a candidate horizon) and the Hawking
temperature $\TH$ calculated for this surface, is infinite, this
surface is a curvature singularity.
\end{itemize}

	Indeed, using e.g. formulae from the book \cite{Wald}, one finds
	for static metrics written in the form (\ref{m1})
	the following expression for the Hawking temperature of a surface
	$u=u^*$ where $\e^{\gamma}=0$, assumed to be a horizon:
\beq
	\TH = \frac{\kappa}{2\pi}, \cm
	\kappa = \lim_{u\to u^*} \ \\e^{\gamma-\alpha}|\gamma'| \label{th}
\eeq
        where we have put the Boltzmann constant $k_{\rm B}$ and the Planck
	constant $\hbar$ equal to 1.  (The same expression can be obtained
	using other methods, such as Euclidean continuation of the
	metric).

Assume now that the surface $u=u^*$ is a candidate
horizon of the metric (\ref{m1}), so that
$\e^{2\gamma} \to  0$ when $u \to  u^*$.
Assume, in addition, that $\kappa =\infty$, while both functions
$\gamma(u)$ and $\e^{\gamma - \alpha}\gamma'$ are monotonic in some
neighbourhood of $u^*$.  Let us show that then the Kretschmann scalar $K
\to  \infty$ as $u \to u^*$.

It is sufficient to prove that $K_1 \to \infty$.

Let us use the fact that $K_1$ (as well as other $K_i$)
and the expression $\e^{\gamma-\alpha}|\gamma'|$ are
unaffected by reparametrizations of the radial coordinate $u$.
With this invariance, any coordinate conditon for $u$ may be chosen without
loss of generality. Let us choose the following one:
\[
     \gamma + \alpha = 0 \, .       \label{11}
\]
Then
\[
K_1 = - \frac{1}{2}[2\gamma'\e^{2\gamma}]' = - \frac{1}{2}[\e^{2\gamma}]'' .
\]
By our assumptions, with (\ref{11}) we have $\e^{2\gamma} \to  0$ and
$(\e^{2\gamma})' \to  \infty$ as $u \to  u^*$.

Let us denote $g(u) = \e^{2\gamma}$, ${1}/{g'(u)} = G(g)$.
Then $G(g) \to  0$ as $g \to 0$.
On the other hand, one can write:
\[
\frac{dg}{du} = \frac{1}{G(g)} \quad\Rightarrow \quad u = \int G(g)dg.
\]
This integral is evidently finite, hence $u^*$ is finite in the coordinates
(\ref{11}). Thus, for a finite value of $u$, we have
$g'= {dg}/{du} \to  \infty$, therefore
\[
 g'' \to  \infty \quad \ \then \ \quad   |K_1| \to  \infty,
\]
which proves the statement.


\section{Criteria for black hole selection}

Black hole (BH) solutions with the metric (\ref{m1}) are conventionally
selected by the following criteria: at some surface $u=u^*$ (horizon)
\begin{description}
\item [C1.]  $\e^\gamma$ $\to  0$ (the timelike Killing vector becomes
              null);
\item [C2.]  $\e^\beta$ is finite (finite horizon area);
\item [C3.]  the integral $t^* = \int \e^{\alpha - \gamma} \to  \infty$
             as $u \to  u^*$ (invisibility of the horizon for an
	     observer at rest).
\end{description}
      The evident requirement that a horizon must be regular (otherwise we
      deal with a singularity rather than a horizon) creates two more
      criteria:
\begin{description}
\item [C4.]  The Hawking temperature $\TH$ is finite;
\item [C5.]  The Kretschmann scalar $K$ is finite at $u=u^*$.
\end{description}

As we have seen in \sect 2, the condition C4 is necessary, but not
sufficient for regularity, and this will be illustrated by the examples
treated below. As for C5, the scalar $K$, due to its structure, is the most
reliable probe for space-time regularity.

The condition C2 is apparently less essential than the others.
In principle, C2 can be cancelled, leading to
a generalized notion of a BH, that with a horizon having an
infinite area, as described in \cite{lousto}.
We will call the BHs satisfying all the criteria C1--C5 type A \bhs,
and those with an infinite horizon --- Type B \bhs.

We shall see that in the general scalar-tensor theory (STT), and, in
particular, in the Brans-Dicke theory all non-Schwarzschild black holes are
type B, while the configurations satisfying C1--C3 turn out to be singular.


\section{Possible Black Holes in the General Scalar-Tensor Theory}

A general Lagrangian describing the interaction between gravity and
a scalar field in four dimensions can be written as
\beq                                               \label{L1}
{\it L} = \sqrt{-g}\biggr(f(\phi)R
         + \frac{\omega(\phi)}{\phi}\phi_{;\rho}\phi^{;\rho}\biggl),
\eeq
where $f(\phi)$ and $\omega(\phi)$ are, in principle, arbitrary
functions of the scalar field $\phi$ (the so-called Bergmann-Wagoner class
of STT). Reparametrization of $\phi$ makes it
possible to deal with one function, for instance, the most conventional
formulation uses $f=\phi$. In what follows it is used here as well.

Performing the conformal mapping
\beq                                                  \label{ct}
g_{\mu\nu} = \phi^{-1}\bar g_{\mu\nu}
\eeq
and omitting a total divergence, we obtain the Lagrangian
\beq                                                  \label{L2}
{\it \bar L}
= \sqrt{-\bar g}\biggr(\bar R + F(\phi)\phi_{;\rho}\phi^{;\rho}\biggl)
\eeq
where
\beq                                                  \label{f1}
F(\phi) = \frac{\omega + 3/2}{\phi^2}
\eeq

The general static, \sph scalar-vacuum
solution for the theory (\ref{L1}) is given by \cite{73,k1}
\bearr                                                    \label{s1}
         ds^2 = \frac{1}{f}\biggl\{
               \e^{-2bu}dt^2 - \frac{\e^{2bu}}{s^2(k,u)}
       \biggr[\frac{du^2}{s^2(k,u)} + d\Omega^2\biggl]
	     \biggr\},                  \\  \lal           \label{s3}
F(\phi)\biggr(\frac{d\phi}{du}\biggl)^2 = S = \const,
\ear
where the last expression is an integral of the scalar and
gravitational field equations. The function $s(k,u)$ is defined as follows:
\beq                                                     \label{f2}
s(k,u) = \vars     {
                    k^{-1}\sinh ku, \ & k > 0 \\
                                  u, \ & k = 0 \\
                    k^{-1}\sin ku,\    & k < 0.  }
\eeq
The constants $b$, $k$ and $S$ are connected by the relation
\beq                                                      \label{r1}
            2k^2\sign k = 2b^2 + S
\eeq
which also follows from the gravitational equations.
The constant $S$ plays the role of a scalar charge.

For $k > 0$, after the coordinate transformation
\beq                                                      \label{t}
	\e^{-2ku} = 1 - \frac{2k}{r} \equiv P(r)
\eeq
the metric can be rewritten as
\beq                                                      \label{m2}
ds^2 = \frac{1}{\phi}\biggr[P^{a }dt^2
          - P^{-a }dr^2 - P^{1-a }r^2d\Omega^2\biggl],
\eeq
with the constants obeying the relation
\beq                                                      \label{c1}
     S = 2k^2(1 -a ^2),\cm      a  = b/k.
\eeq

Let us analyze the possible existence of BHs in the general STT,
i.e. with variable $\omega=\omega(\phi)$. As can be seen from
(\ref{s3}), $S < 0$ corresponds to anomalous theories ($\omega +
3/2 < 0$), having a negative kinetic term in the Lagrangian (\ref{L2}),
while for $S = 0$ we have $\phi = \const$, i.e. general relativity.

From the viewpoint of Criteria C1--C3,
there are four opportunities for BH existence \cite{k1}:
\begin{enumerate}
\item $k > 0$, $S<0$, $u^* \to  \infty$;
\item $k > 0$, $S>0$, $u \to  \infty$ is a regular sphere and a horizon may
    be found beyond it by proper continuation (example: a BH with a
    conformal scalar field);
\item $k = 0$, $S<0$, $u^* \to  \infty$;
\item $k < 0$, $S<0$, $u^* = \pi/|k|$.
\end{enumerate}

     Let us consider each case separately, except for the second one, since
     it is hard to handle in a general form due to the continuation.  We
     will try first to apply the requirements C1--C3.  One can notice that
     in all cases to be considered the theory is anomalous%
\footnote{As shown in \cite{k1}, the cases when the sphere $u=\infty$ is
     regular and admits an extension of the static coordinate chart, are
     very rare, and even when it is the case, such configurations turn out
     to be unstable due to blowing-up of the effective gravitational
     coupling \cite{78}.}.

\begin{enumerate}
\item $k > 0$. In this case we can use the metric (\ref{m2}) with
      (\ref{c1}). As $P \to  0$, due to C2 the scalar field behaves like
\beq
      \phi \sim   P^{1 - a }
\eeq
while from C3 it follows $a  > 1$.
The Hawking temperature is calculated from (\ref{th}) and the asymptotic
form of $\omega(\phi)$ may be found from (\ref{s3}). The result is
\beq
\TH \sim \lim_{P\to 0} P^{a -1} = 0, \cm
        \omega + \frac{3}{2} \to  -\frac{1}{2}\frac{a +1}{a -1}.
\eeq
However, the term $K_1$ of the Kretschmann scalar behaves as $P^{-1}$ and
tends to infinity as $P\to 0$.

\item
$k = 0$. Using the metric (\ref{s1}) with $k = 0$, the
conditions C1--C3 lead to $u^*=\infty$, and as $u\to \infty$,
\beq
    \phi \sim \frac{1}{u^2}\e^{2bu}, \cm b >0,
\eeq
For $\TH$, $\omega$ and $K_1$ we obtain:
\beq
\TH  \sim \lim_{u \to  \infty} u^2\e^{-2bu} = 0, \cm \omega \to - 2; \cm
	K_1 \sim 4b^2 u^2 \to \infty.
\eeq
\item
$k<0$. A possible horizon is at $u^* = \pi/|k|$,
where $\phi \sim 1/\Delta u^2$, $\Delta u \equiv |u - u^*| $.
A calculation gives:
\bearr
| \e^{\gamma-\alpha}\gamma'|  \sim \Delta u \to  0
       \quad  \then \quad   \TH = 0; \nnn
	\frac{\phi_u}{\phi} \sim \frac{1}{\Delta u} \to  \infty
	    \ \then \ \omega + \frac{3}{2} \to  -0  .
\ear
The scalar $K$ is in this case finite; the behaviour of $g_{00}$ and
$g_{11}$ near the horizon is similar to that in the extreme
Reissner-Nordstr\"om solution.
\end{enumerate}

We notice that in all these cases the Hawking
temperature calculated for the assumed horizons is zero.
However, $\TH < \infty $ is only necessary but not sufficient for
regularity. In the above cases of the general STT,
the Kretschmann scalar turns out to be infinite and reveals
a singularity in cases 1 and 3, while $K < \infty$ only in case 4 ($k<0$).
Thus for $k\geq 0$ the surfaces of finite area, satisfying the conventional
criteria of an event horizon, turn out to be singular, and only Type B
\bhs\ can exist.  This will be demonstrated explicitly for the Brans-Dicke
theory.

\section{Brans-Dicke Black Holes}

Let us now analyze the four-dimensional spherically symmetric solution in
Brans-Dicke theory ($\omega = \const$). In this case we can write
explicitly $\phi=\phi_0\e^{su}$, where $\phi_0$ and $s$ are constants and
$s^2 = S/(\omega+3/2)$. The solution (\ref{s1}), (\ref{s3}) for
different $k$ then coincides (up to reparametrization) with different
classes of the Brans solution \cite{Brans} to the vacuum Brans-Dicke
equations.

Let us again consider the cases $k > 0$, $k = 0$ and $k <0$ separately.

\subsection{$k > 0$}

Using, as before, the transition (\ref{t}) and
denoting $\xi=s/2k$, we can write the metric in the form (\ref{m2}), namely,
\beq                                                         \label{bds}
	ds^2 = P^{-\xi}\Bigl(P^{a }dt^2 - P^{-a }dr^2
                 - P^{1 - a }r^2 d\Omega^2 \Bigl),
\eeq
and the constants are related by
\beq
	(2\omega+3) \xi^2 = 1-a ^2.                         \label{xi}
\eeq

The limit $u\to\infty$ corresponds to $r\to 2k$, or $P\to 0$.
The condition C1 ($g_{00} \to  0$ as $P \to  0$) implies
$a -\xi >0$. On the other hand, the condition C5 with $\xi\ne 0$
leads to $a  + \xi \geq 2$. Indeed, we have
\beq
	K_1 \sim (1 - a)(a -\xi)P^{a +\xi-2} .       \label{kr+}
\eeq
With (\ref{xi}) this implies, in particular, $2\omega + 3 < 0$
(anomalous theory). Combining the two inequalities for $a $ and $\xi$, we
obtain the following allowed range for these constants:
\beq
	a >1, \inch     a  > \xi \geq 2- a .             \label{range+}
\eeq

A finite limit of $g_{22}$ (Criterion C2) as $P\to 0$ is obtained if
$a +\xi=1$, and the only nonsingular case ($K<\infty$) under this condition
is $\xi=0$, $a =1$ --- the case when the Brans-Dicke theory reduces to
general relativity and the solution is just the Schwarzschild metric.
(In what follows we exclude it from consideration.)
Moreover, the solution with $a +\xi=1$ violates Criterion C3 when $\xi>0$
since  $t^* \sim \int P^{\xi - 1}dr$, which converges in this case.
Thus the non-Schwarzschild surface $ P = 0$ is not only singular, but
visible from outside for an observer at rest if $\xi > 0$, i.e.,
$\omega >- 2$.

Let us now calculate the Hawking temperature for the solution
(\ref{bds}). We have $\e^{\gamma - \alpha} \sim P^{a }$,
$\gamma'\sim P^{-1}$ and consequently
\beq
    \TH \sim P^{a  - 1}\Big|_{P\to 0} = 0
\eeq
in the allowed range (\ref{range+}).

In the above case of finite limiting surface area ($a +\xi=1$) one has
$T_H= \infty$ if $a <1$, $\xi>0$ (this is just a visible naked
singularity), but $T_H=0$ if $a >1$. But, as we remember,
the Kretschmann scalar is infinite if $a +\xi <2$. This is a clear
illustration of the fact that a finite and even zero Hawking
temperature does not guarantee a non-singular horizon.

In the allowed range (\ref{range+}) Criteria C1, C3--C5 hold, but the
horizon radius is infinite and $\TH=0$ --- a Type B \bh. Let us try, for
better understanding of its properties, to perform a Kruskal-like extension
of this solution beyond the horizon, i.e., to find a coordinate chart
without an apparent metric singlularity $g_{00}=0$ at $r=2k$.
To this end, let us first introduce the null coordinates $v$ and $w$
\beq                                 \label{vw}
	v = t + x, \inch
	w = t - x
\eeq
    where
\beq                                 \label{r*}
        x = \int  dr/P^{a },
\eeq
    so that $x\to\infty$ as $r\to\infty$ and $x\to -\infty$ as $r\to 2k$.
    The metric (\ref{bds}) takes the form
\beq                                                        \label{ds-vw}
        ds^2 = P^{a -\xi} dv dw - P^{1-a -\xi }r^2 d\Omega^2.
\eeq
    The integral (\ref{r*})
    admits a closed expression only for integer $a $:
\beq                                         \label{sr*}
	x = r + 2ka \ln(r-2k)
		+ \sum_{n=1}^{a -1}\frac{(-2k)^{n+1}}{n}
			{a  \choose n+1} \frac{1}{(2k-r)^n}.
\eeq
    In the general case, the asymtotic behaviour of $x$ as $r\to 2k$ is
\beq
	x \sim (r-2k)^{1-a }, \inch   a  > 1.             \label{x-as}
\eeq

    Our next step must be a transition to some coordinates $V=V(v)$ and
    $W=W(w)$ eliminating the zero limit of $g_{vw}$
    in (\ref{ds-vw}) as $x\to -\infty$, or,
    equivalently, as $v \to -\infty$ and/or $w\to \infty$.

    Let us look for the new coordinates in the following asymptotic form:
\beq
	v \sim V^p,    \inch        w \sim W^p               \label{vV}
\eeq
    with some constant $p$ to be determined from the regularity
    requirement (it is the same for $v$ and $w$ from symmetry
    considerations).  Assuming $v\to -\infty$ and finite $w$, we find the
    asymptotic form of the $(U,V)$ part of the metric as
\beq
    (r-2k)^{1-\xi+(1-a )/p}dV\,dW,                          \label{VW}
\eeq
    which provides regularity at the horizon if
\beq
    p = (1-a )/(1-\xi),                                     \label{p}
\eeq
    provided $\xi\ne 1$. The same expression for $p$ is found if we
    consider the limit $w \to \infty$ keeping $v$ finite.

    With (\ref{p}), $V$ is asymptotically related to $r-2k$ by
\beq
    V \sim (r-2k)^{1-\xi} + \const.
\eeq
    Hence for $\xi > 1$, $V \to \infty$ as $v \to -\infty$, while for
    $\xi < 1$ the coordinate $V$ has a finite limit at the
    horizon. The same applies to the coordinate $W$.
    We see that the allowed range of $\xi$ (see (\ref{range+}))
    is divided into two subdomains by the line $\xi  = 1$.

    If $\xi=1$, the assumption (\ref{vV}) does not work and the regularity
    is achieved using the transformation
\beq
    V \sim \ln |v|, \cm  W \sim \ln |w|.                     \label{vV'}
\eeq
    The coordinates $V$ and $W$ tend to infinity as $x\to -\infty$.

    Thus for $\xi < 1$, we obtain a more or less common picture of a BH,
    where particles can arrive at the horizon in a finite proper time, and
    in principle they can cross it, entering the BH interior.  Let us call
    such configurations Type B1 \bhs.

    When $\xi\geq 1$, the metric (\ref{VW}) behaves asymptotically as
    $dV\,dW$ with an infinite range of $V$ and $W$, i.e., the assumed
    horizon is infinitely far and it would take an infinite time for any
    particle to reach it. Recalling that, in the same limit, $g_{22}\to
    \infty$, this configuration (to be called a Type B2 \bh) resembles a
    wormhole, although, in general, the correct flat-space asymptotic
    conditions are here not observed.

    This interpretation is confirmed by a direct study of radial
    geodesics. The equation for a radial timelike trajectory in a static
    space-time with the metric (\ref{m1}) can be written, after first
    integration, as
\beq                                                     \label{geod}
    \biggr(\frac{dr}{d\tau}\biggl)^2 =
               \e^{-2\alpha} (E^2 \e^{-2\gamma} -1),
\eeq
    where $E$ is an integration constant (energy) and $\tau$ is the proper
    time. Close to a horizon ($\e^{\gamma}\to 0$)
    the second term can be neglected as compared with
    the first one. In our case $\e^{-\alpha-\gamma}\sim P^{2\xi}$,
    and leads to
\beq
    d\tau \sim (r-2k)^{-\xi}.                   \label{tau+}
\eeq
    Hence a particle needs an infinite proper time to reach $r=2k$
    when $\xi \geq 1$ and a finite proper time when $\xi < 1$,
    in full agreement with the previous analysis of the metric behaviour.

    One can notice that, under fixed $a$, large values of
    $|\omega|$, necessary to conform with present-day observations,
    correspond to small $\xi$, i.e. to Type B1 BHs with horizons available
    in finite proper times.

\subsection{$k = 0$}

    In this case the metric (\ref{s1}) takes the form
\beq
ds^2 = \e^{-su}\biggr[\e^{-2bu}dt^2 - \frac{\e^{2b u}}{u^2}\biggr(
	\frac{du^2}{u^2} + d\Omega^2\biggl)\biggl], \cm
       	  s^2(\omega + 3/2) = -2b^2.                            \label{ds0}
\eeq
    There is no case of finite $g_{22}$ as $u \to  \infty$. One has
\bear
   \gamma'\e^{\gamma-\alpha} \eql - u^2(b + s/2)\e^{-2bu},\\
      K_1  \al\sim\al b(b + s/2)u^4\e^{(s-2b)u}\, .
\ear
    Thus $K_1 \to  \infty$ if the radius $r = \sqrt{|g_{22}| }
        \to  0$ and $K_1 \to 0$ if
    $r \to  \infty$ as $u \to  \infty$. The case $b = 0$ is
    trivial, while $b + {s}/{2} = 0$ corresponds to a ``force-free"
    ($g_{00} = \const$) solution with a special value $\omega = -2$.
    The allowed range of the integration constants, where $g_{00}\to 0$ and
    $K_1 <\infty$, is
\beq
	b > 0, \inch   2b > s > -2b.                \label{range0}
\eeq
    In this range the other $K_i < \infty$ as well.
    The Hawking temperature for these solutions is again zero and
    $|2\omega + 3|  > 1$.
    These solutions, regular as $u\to \infty$, describe Type B \bhs.

    It is hard to obtain an explicit (even asymptotic) expression for
    a coordinate transformation needed for a Kruskal-like extension,
    since now $x = \int du\cdot \e^{2bu}/u^2$.
    We can, however, distinguish B1 and B2 BHs, using, as
    before, the geodesic equation (\ref{geod}). One has
\beq
	d\tau \sim u^{-2}\e^{-su} du,                     \label{tau0}
\eeq
    so that $\tau$ is finite for $s > 0$ and infinite for $s<0$
    ($s=0$ is excluded since leads to general relativity). Thus the
    range (\ref{range0}) is again divided into two halves: for $s>0$ we
    deal with a Type B1 BH, for $s<0$ - with that of Type B2.

    According to (\ref{ds0}), large values of $|\omega|$ correspond to
    small $s$, which may be, however, of either sign, so that both Type B1
    and B2 BHs can in principle be compatible with observations.

\subsection{$k < 0$}

    Now the solution (\ref{s1}) is
\beq
	ds^2 = \e^{-su}\biggr[\e^{-2bu}dt^2
	     - \frac{k^2\e^{2bu}}{\sin^2{ku}}\biggr(
	\frac{k^2du^2}{\sin^2{ku}} + d\Omega^2\biggl)\biggl],
\cm
s^2(\omega + \frac{3}{2}) = - k^2 - 2b^2.
\eeq
    This is a wormhole solution \cite{73}, with two flat asymptotics at
    $u = 0$ and $u = u^* = \frac{\pi}{|  k| }$. Due to the monotony of
    $g_{00}$, the masses at $u = 0$ and $u = u^*$ have opposite signs;
    there is neither a singularity, nor a horizon.

\section{Concluding remarks}

    We can make the following general observations from the above analysis:
\begin{enumerate}
\item
    Black holes exist in anomalous scalar-tensor theories,
    i.e., when the kinetic term of the scalar field is
    negative, contrary to some statements put forward in \cite{kim}.
\item
    For $k\geq 0$, no conventional (Type A) BHs can exist.  BHs
    with an infinite area (Type B) do exist, as confirmed explicitly for a
    special case --- the Brans-Dicke theory.  It is shown that they in turn
    split into two classes, B1 and B2, with, respectively, finite and
    infinite proper time needed for an infalling particle to reach a
    horizon.
\item
    In the case $k<0$ Type A \bhs\ can exist, but only in theories with
    variable $\omega$, and such explicit examples are yet to be found.
\item
    Type B2 BHs are obtained only in cases when
    the Brans-Dicke scalar field $\phi\to 0$ at the horizon, i.e., when the
    effective gravitational coupling tends to infinity.
\end{enumerate}

    Trying to perform an explicit Kruskal-like extension beyond the regular
    horizons (which makes sense for Type B1 BHs, available for particles),
    one meets, in the general case, negative quantities in fractional
    powers, and only in special cases when those powers are integer, the
    extension is performed using the well-known methods. The situation
    needs further study and we hope to return to it in future works.

{\bf Acknowledgements} We thank George Matsas for many helpful discussions. This work was partially supported by CNPq (Brazil) and CAPES (Brazil).


\begin{thebibliography}{90}

\bibitem{lousto}
M. Campanelli and C.O. Lousto, {\it Int. J. Mod. Phys.}
               {\bf D2}, 451 (1993).

\bibitem{kim} H. Kim and Y. Kim, {\it Nuovo Cim.} {\bf 112B}, 329 (1997).

\bibitem{73}
K.A. Bronnikov, {\it Acta Phys. Polon.} {\bf B4}, 251 (1973).

\bibitem{k1}
K.A. Bronnikov, {\it Grav.\& Cosm.} {\bf 2}, 221 (1996).

\bibitem{Wald}
R. Wald, ``General Relativity", Univ. of Chicago Press, Chicago, 1984.

\bibitem{78}
K.A. Bronnikov and Yu.N. Kireyev, {\it Phys. Lett.\/} {\bf A 67}, 95 (1978).

\bibitem{Brans}
C. Brans, {\it Phys. Rev.} {\bf 125}, 2194 (1962).

\end{thebibliography}
\end{document}